\documentclass[preprint,rmp,sort&compress,authoryear]{revtex4}
\usepackage{graphicx}
\begin{document}
\title{The central role of line tension in the fusion of biological
membranes}
\author{M. \ Schick}
\affiliation{Department of Physics, University of Washington,  Box
  351560, Seattle, WA 98195-1560}  
\author{K. \ Katsov} 
\affiliation{Materials Research
Laboratory, University of
California, Santa Barbara CA 93106-5080}
\author{M. M{\"u}ller} 
\affiliation{Department of Physics, University of Wisconsin-Madison,
  Madison, WI 53706-1390}
\date{\today}

\begin{abstract}
 Recent progress in the fusion of biological membranes is reviewed to
highlight the central role played by the line tension,  which permits  exquisite
control of the process.
\end{abstract}

\maketitle

{\em La plus expresse marque de la sagesse, c'est une esjou{\"i}ssance
constante; son etat est comme des choses au dessus de la Lune; toujours
serein.
\vskip 0.2truein
\hskip 4 truein Michel de Montaigne}

The importance of the fusion of biological membranes is sufficiently clear 
that only a few keywords, such as endocytosis, intracellular trafficking, 
synaptic release, and viral entry, should suffice to remind the reader of 
it. For all its importance, the physics of this topological rearrangement 
is not well understood at all. In fact, what we see as the central 
conundrum which fusion presents seems, with notable exceptions
\cite{chernomordik03}, not to have been addressed explicitly. That 
conundrum is the following. In order for any vesicle to be useful, it must 
be relatively stable. In particular, its enclosing membrane must be stable 
to the occurrence of long-lived holes which are thermally activated. Yet 
in order to undergo fusion, just such long-lived holes must occur at
some point along the fusion pathway. 
It would seem that 
vesicles could {\em either} be stable, {\em or} they could undergo fusion, 
but not both. How they actually manage to exhibit these two conflicting properties
is the conundrum. Because of 
recent work on this problem, some to be published elsewhere 
\cite{katsov05}, we believe we understand the resolution of this puzzle. 
Because line tension is at the heart of this resolution, we thought it a 
very appropriate subject to be included in a volume honoring Ben Widom 
whose interest in, and explication of, this concept is long-standing 
\cite{rowlinson82}.

Let us briefly review the situation. We begin with two membranes, each 
consisting of two layers of amphiphiles, or lipids. In general the 
head groups of the lipids like to be surrounded by water. To bring the 
membranes sufficiently close together so that fusion can occur, the 
interposed water must be removed, at least in some region between the 
membranes. To remove this water takes energy which, presumably, is provided {\em in 
vivo}, by fusion proteins. Due to the loss of water between membranes, the 
free energy per unit area, or surface tension, of the membranes increases. 
One possible response of the system to this increase is to undergo fusion 
because this process, by making holes in the membranes, decreases the 
membrane area and thus the free energy. The canonical way this has been 
thought to occur (see \cite{chernomordik03} and references therein) 
was first suggested by Kozlov and Markin \cite{kozlov83}, 
and is illustrated in Fig. 1.

\begin{figure}
\includegraphics[scale=.5]{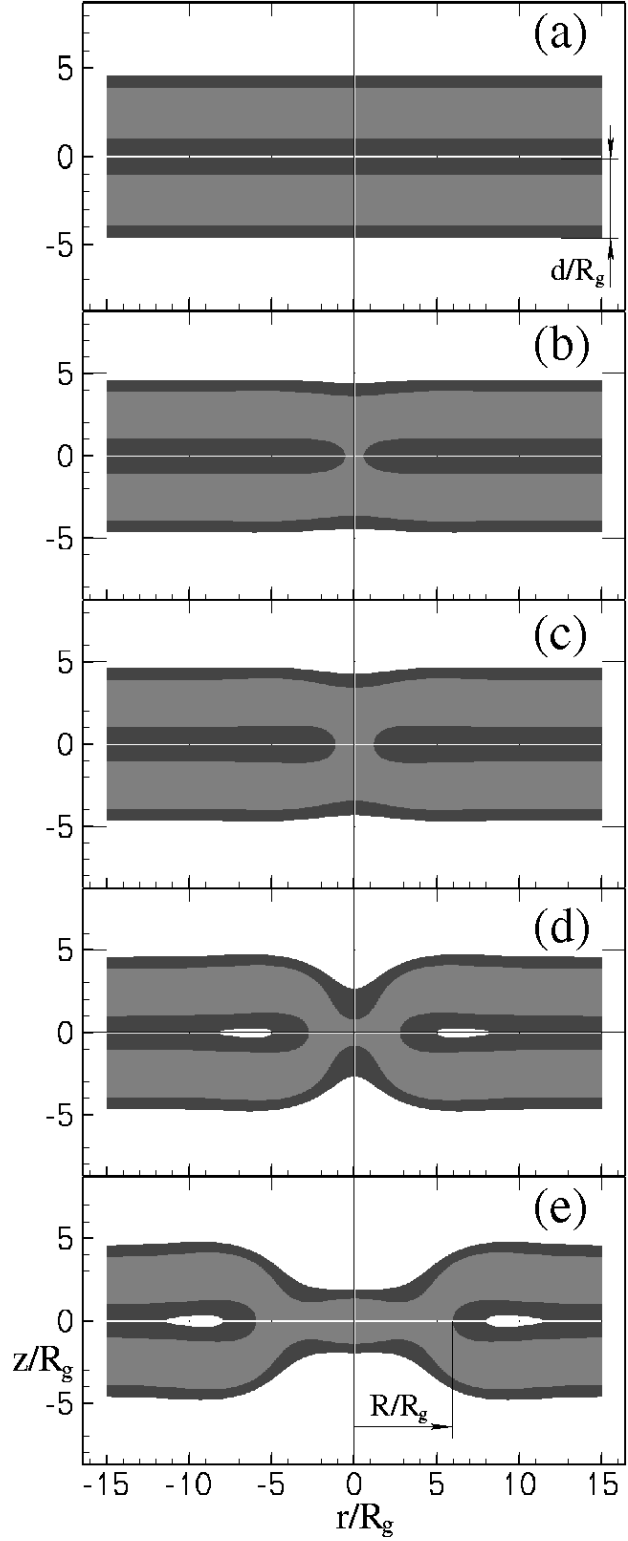}
\caption{Density profiles of structures from bilayers in apposition,
(a), passing through a metastable stalk (c), to a hemifusion diaphragm
(e). Figures are shown in the $r,z$ 
plane of cylindrical coordinates.  } 
\label{prof-stalk}
\end{figure}

In panel a) we see two bilayers under zero tension, They are composed of 
amphiphiles, block copolymers in this case, which contain a fraction 
$f=0.35$ of the hydrophilic component.  Only the majority component is 
shown at each point: solvent segments are white, hydrophilic and 
hydrophobic segments of the amphiphile are dark and light correspondingly. 
Distances are measured in units of the polymer radius of gyration, $R_g$, 
which is the same for both the amphiphiles and for the homopolymer 
solvent. 
In (b), tails of some amphiphiles in a small region have 
turned over, attempting to form an axially-symmetric ``stalk''. This
panel shows 
the transition state to the formation of the stalk, and  
panel (c) shows the metastable stalk itself. 
After the stalk forms, the layers pinch down and expand, pass 
through a second intermediate, shown in (d), and arrive at a hemifusion 
diaphragm, (e).  A hole then forms in this 
diaphragm, which completes formation of the fusion pore. Note that the 
conundrum is not addressed explicitly by this scenario.  
However one can observe that this mechanism requires a hole to form only 
in the {\em one} hemifusion diaphragm rather than in the {\em two} 
bilayers separately.

\begin{figure}
\includegraphics[scale=0.37]{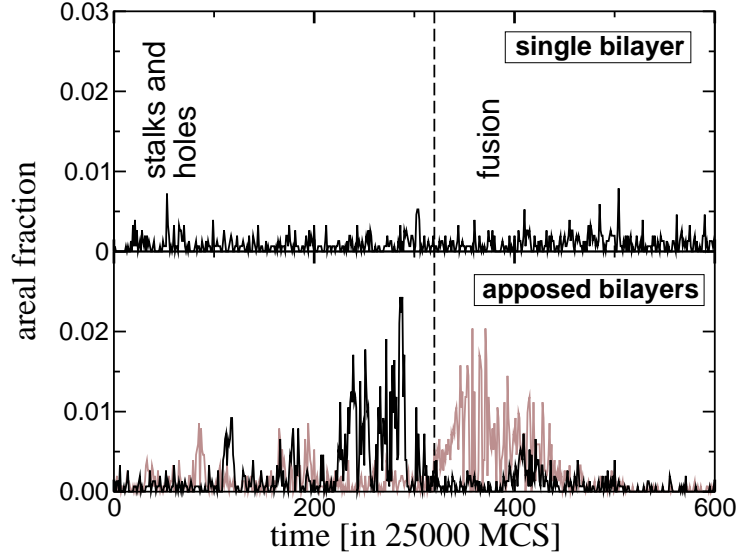}
\caption{Area of holes vs. time in the system of two apposed bilayers,
bottom, and in an isolated bilayer, top.}
\label{holetime}
\end{figure}

Sometime ago, we decided to watch, via Monte Carlo simulation, the fusion 
process unfold in a system of bilayers formed by block copolymers in 
homopolymer solvent \cite{mueller02}. Our choice of this system of 
non-biological amphiphiles was motivated by the fact that we had experience in 
simulating such amphiphilic copolymers, and our belief, as physicists, 
that the fusion process was probably universal. The time and energy scales 
would vary from system to system, but not the pathway of the process 
itself. Furthermore vesicles of block copolymer form a novel family which 
is currently being investigated for its technological possibilities 
\cite {discher99}. Details of the simulation can be found in 
\cite{mueller02} and \cite{mueller03}, but the results can be summarized 
as follows. Upon putting the bilayers under tension and in close apposition, 
we did see the formation of an axially symmetric stalk. We expected to see 
the stalk expand radially, but it did not. Instead, it expanded asymmetrically, 
forming a worm-like structure which moved about. We also observed that 
once the stalk formed, the rate of hole formation in either bilayer rose 
dramatically. This is shown in Fig. \ref{holetime} where, in the lower 
panel, the rate of hole formation in each bilayer, one in black, the other 
in gray, is seen to rise dramatically after about 200 time steps when we 
know, independently, that a stalk had formed. The rate of hole formation 
in a {\em single} bilayer is shown in the upper panel for comparison.

\begin{figure}
\includegraphics[scale=0.39]{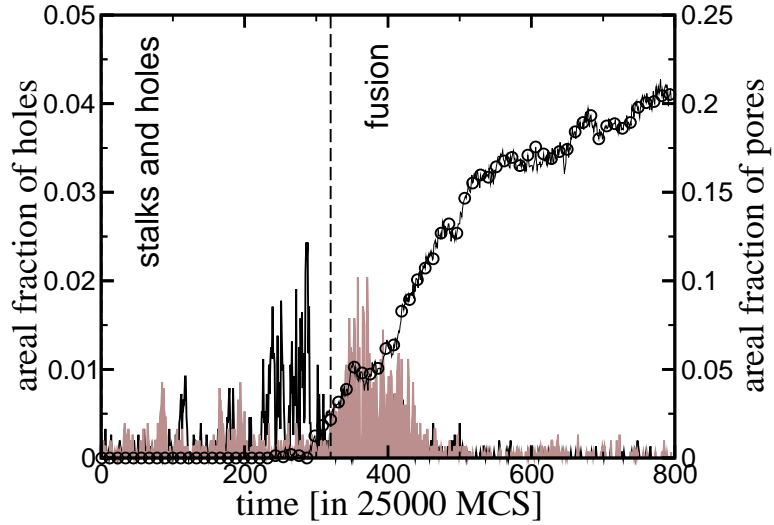}
\caption{Area of pore (symbols) and of holes (lines) for one simulation
run, the same as shown in Fig. \ref{holetime}. Note the different scale
for pore and hole areas.} 
\label{poretime}
\end{figure}

Furthermore, we could determine that the stalk and the newly-created holes 
were correlated; that is, for the most part, the holes formed very near to 
the stalk. Once a hole formed in one bilayer next to the stalk, the 
latter, which we had observed to be quite mobile, proceeded to walk around 
the hole, thereby forming something like a hemifusion diaphragm. Once a 
second hole pierced this diaphragm, the fusion pore was complete. In a 
slightly different scenario, we saw a hole form in one bilayer, and the 
stalk begin to walk around it. Before it completely surrounded the first
hole, a 
second one appeared in the other bilayer near the stalk. The stalk then 
had to corral the two holes, walking around them both, to complete the 
fusion pore.

\begin{figure}
\includegraphics[scale=0.5]{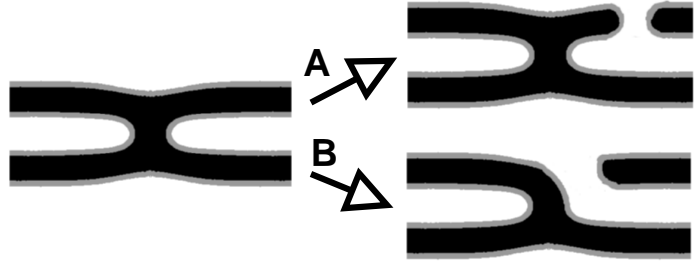}
\caption{A schematic diagram which makes plausible that the line tension
of a hole which forms near a stalk, as in path B, is less than if it
forms far from a stalk, as in A.  } 
\label{schematic}
\end{figure}

It is clear that in the mechanism we saw, the formation of a fusion pore 
is closely correlated, in space and time, with hole creation. This 
correlation is seen in Fig. \ref{poretime}; the formation of pores closely 
follows in time the onset of hole formation triggered somehow by stalk 
formation . There is a clear experimental differentiation between this new 
mechanism, and the standard hemifusion mechanism discussed earlier. This 
consequence, transient leakage, can be understood from Fig. 
\ref{schematic} which shows that for a certain period of time, there is a 
hole from at least one of the vesicles to the outside during the fusion 
process. How much leakage there is depends on what molecule one is 
observing, as each will have its own characteristic time to diffuse 
through the hole. If this 
time is significantly greater than the time for the stalk to surround the 
hole and seal it up, there will be little, if any, observable leakage. 
However if the
time to diffuse to the hole is much less than the sealing time, there
will be. Just such leakage, 
correlated with fusion in the manner of Fig. \ref{poretime}, was recently 
observed in an elegant experiment \cite{frolov03}.

How do we understand the behavior we have seen in our model, and by others 
\cite{noguchi01} in a more simplified model? We had an idea as to what was 
going on, and to verify it, we embarked on a series of self-consistent 
field calculations \cite{katsov04,katsov05} on the same system as had been 
simulated, from which various free energies could be calculated 
explicitly.

\begin{figure}
\includegraphics[scale=1.]{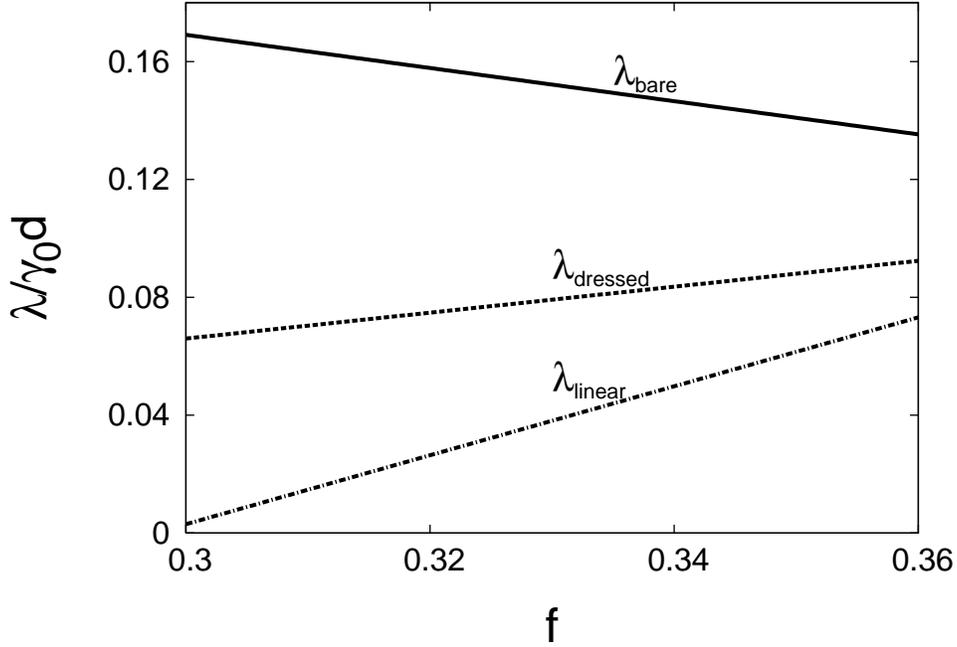}
\caption{Line tensions of a linear stalk, $\lambda_{linear}$, of a bare
hole in a membrane, $\lambda_{bare}$, and of a hole which forms next to
a stalk, $\lambda_{dressed}$. All line tensions are in units of $\gamma_0
d$, the bare hydrophilic-hydrophobic
surface tension multiplied by the bilayer thickness.} 
\label{linetensions}
\end{figure}

First of all, we can understand the wandering of the elongated stalk by 
calculating its free energy per unit length, that is, its line tension. Not 
surprisingly, we find that it varies with the architecture of the block 
copolymer, which is described by the fraction, $f$, of hydrophilic 
monomers in the diblock. For values of $f$ in the vicinity of 0.5, the 
system makes bilayers. As $f$ is reduced, the majority hydrophobic 
component wants more space to explore more configurations. Eventually
this will cause the 
system to undergo a phase transition to an inverted hexagonal 
phase consisting of cylinders with the minority hydrophilic parts confined 
to the smaller region inside the cylinders and the majority hydrophobic
part filling the larger region outside. 
These cylinders are separated by structures that look very much like stalks 
that are stretched out parallel to them. The phase transition occurs at a value 
of $f\approx 0.31.$ The line tension, $\lambda_{linear}$, of a linear stalk 
which we have calculated is shown in Fig. \ref{linetensions}.
It is given there in units of $\gamma_0 d$, where $\gamma_0$ is the
surface free energy per unit area
between 
coexisting regions of hydrophobic and hydrophilic molecules, and $d$
is the thickness of 
a bilayer. 
We note that we have found \cite{katsov04} successful fusion 
can occur over an interval of architectures from about $0.31<f<0.35$. 
In this interval the 
line tension of the linear stalk does not exceed 0.06, and becomes extremely 
small as $f$ approaches the transition to the inverted hexagonal phase. 
This is, of course, no coincidence, and tells us that this fusion 
mechanism becomes more favorable as the architecture of the lipids in the 
system become more like those of ``hexagonal formers''. The upshot is
that, as this transition is approached, it costs very little energy for
the stalk to walk around as we had observed.

Secondly it seems intuitively clear from Fig. \ref{schematic} that the 
line tension of a hole when it forms near a stalk is less than that of a 
hole formed far from a stalk. To verify this intuition, we have calculated 
the line tension, $\lambda_{bare}$, of a ``bare'' hole, that is, the line
tension associated with the formation of a linear edge of a bilayer
membrane. We have also calculated the line tension, $\lambda_{dressed}$,
of a hole in one
bilayer which is next to a linear stalk; that is, the line tension
associated with the linear defect at which two bilayers join to form one
bilayer, as in Fig. \ref{schematic}. These quantities are also shown in
Fig \ref{linetensions}. We note that the line tension of a hole formed
next to a stalk is indeed
lower than that of a bare hole. Depending upon the architecture, the
reduction in line tension is at least a factor of two. Further, the
dependence on architecture is such that the reduction is greater the
smaller the value of $f$, that is, the more hexagonal-forming the
amphiphiles are. 

We can now understand the conundrum posed by fusion. Let us consider the
simple approximate phenomenolgical 
expression for the free energy of a hole of radius
$R$ and line tension $\lambda_{bare}$ in a
membrane of surface tension $\gamma$, which generally is much smaller
than $\gamma_0$,
\begin{equation}
F_{bare}(R)=2\pi R\lambda_{bare}-\pi R^2\gamma.
\end{equation}
In order for the hole to expand, and become long-lived, it must pass the
barrier of free energy 
\begin{equation}
F_{bare}^*(R_{bare}^*)=\pi\lambda_{bare}^2/\gamma
\end{equation} 
which occurs at $R_{bare}^*=\lambda_{bare}/\gamma$.
Stability of a membrane is guaranteed by the fact that the rate
of hole formation, which is proportional to the Boltzmann factor
\begin{equation}
P_{bare}\sim \exp\left\{-\frac{\pi\lambda_{bare}^2}{\gamma k_BT}\right\}
\end{equation}
is small. For example, for the system simulated,
$\lambda_{bare}/\gamma_0d\approx0.15$, $\gamma_0 d^2/kT\approx 60$, and
$\gamma/\gamma_0\approx 0.75$, from which we obtain $P_{bare}\approx
4\times 10^{-3}$. 

Now however, let the hole form next to a stalk which almost completely
surrounds it, reducing its line tension 
from $\lambda_{bare}$ to $\lambda_{dressed}$. The barrier to stable hole
formation is now only
\begin{equation}
F_{dressed}^*(R_{dressed}^*)=\pi\lambda_{dressed}^2/\gamma,
\label{dressed}
\end{equation}
and the relative rate of formation of a long-lived hole compared to what
it was without the stalk is
\begin{equation}
\frac{P_{dressed}}{P_{bare}}=\exp
\left\{\frac{\pi \lambda_{bare}^2}{\gamma k_BT}
\left(1-\frac{\lambda_{dressed}^2}{\lambda_{bare}^2}\right)\right\}
\end{equation}
With $(\lambda_{dressed}/\lambda_{bare})^2$ much less than unity, the rate of
hole formation increases almost by the large factor $P_{bare}^{-1}>>1$.
In particular, taking the reduction in line tension to be a modest 
$\lambda_{dressed}/\lambda_{bare}=1/2$, and
$P_{bare}=4\times 10^{-3}$, we obtain 
\begin{equation}
\frac{P_{dressed}}{P_{bare}}\approx 60, 
\end{equation}
so that the rate of hole formation rises
by over an order of magnitude, in qualitative agreement with the
results from simulation. 

This increase in the rate of hole formation is predicted to be 
far more dramatic for a biological membrane. In such a system,\ 
$\lambda_{bare}\approx 2.6\times 10^{-6}$erg/cm
\cite{moroz97,zhelev93}. To obtain an order of magnitude for the local
tension in a biological membrane undergoing fusion, we consider a
scenario \cite{kozlov98} in which six hemagglutinin molecules
release their energy of conformational change, about 60\ kT per
molecule, within a circular area of radius 4 nm. This yields an
estimate of 30 erg/cm$^2$. For illustration we shall take a $\gamma$ of
10 erg/cm$^2$ which is not unreasonable.
We then find that $P_{bare}\approx 3 \times 10^{-22}$. A single  membrane is stable
indeed! The relative rate of formation of a long-lived hole in the
presence of a stalk compared to the rate without it becomes, for the same
modest reduction in line tension by a factor of two, 
\begin{equation}
\frac{P_{dressed}}{P_{bare}}\sim 1\times 10^{16}\ !
\end{equation}
Of course the exact expression for the free energy of the intermediate
in which the hole is only partially surrounded by the stalk will differ
in detail from the simple expression given in
Eq \ref{dressed}. Nevertheless, the key determinant in this free energy,
the quadratic dependence upon the line tension, will remain.  
It is the fact that {\em fusion is a thermally excited event for
which the rate is proportional to the exponential of the square
of the line tension} which explains the conundrum of fusion. As
long as the line tension is ``normal'', a membrane is {\em extremely} stable
to thermally excited holes. But because the membrane {\em is} so stable, and
because the line tension appears {\em squared}
in the exponent, any mechanism, such as the one we have proposed, which
even slightly affects  the line tension
will greatly affect the rate of hole formation, and therefore the rate of fusion.
Thus does fusion become possible!

\section{Acknowledgments}
For very useful conversations, we thank  L. Chernomordik, F. Cohen,
M. Kozlov, B. Lentz, D. Siegel, and J. Zimmerberg. We are particularly
grateful to V. Frolov for sharing his knowledge and expertise with
us. This material is based upon work supported by the National Science
Foundation under Grant No. 0140500. Additional support was provided by
the DFG  Mu1647/1. Computer time at the NIC J{\"u}lich is  gratefully
acknowledged.

\end{document}